\documentclass[conference]{IEEEtran}

\usepackage{amsmath,amssymb,amsfonts}
\usepackage{graphicx}
\usepackage{booktabs}
\usepackage{hyperref}
\usepackage{xcolor}
\usepackage{multirow}
\usepackage{tabularx}
\usepackage{enumitem}

\hypersetup{
  colorlinks=true,
  linkcolor=blue,
  citecolor=blue,
  urlcolor=blue
}

\begin{document}

\title{Combating Organized Platform Abuse:\\Amplifying Weak Risk Signals with Structural Information}

\author{
  \IEEEauthorblockN{HE Meng}
  \IEEEauthorblockA{
    Trust \& Safety, Grab Holdings\\
    China\\
    he.meng@grabtaxi.com\\
    \textit{Corresponding: hemeng86@gmail.com}
  }
  \and
  \IEEEauthorblockN{Jia Long LOH}
  \IEEEauthorblockA{
    Trust \& Safety, Grab Holdings\\
    Singapore\\
    jialong.loh@grabtaxi.com
  }
}

\maketitle

% ============================================================
\begin{abstract}
Large-scale online service platforms face severe challenges from organized platform abuse---multiple forms such as credit card fraud and promotion abuse continually emerge, characterized by large numbers of involved accounts, rapid outbreaks, and constantly shifting tactics.
Existing mainstream approaches---whether heuristic rules limited in precision, supervised learning with insufficient generalization, or graph models that are engineering-heavy and dependent on seed users---have failed to address such threats effectively.

This paper returns to first principles and, starting from the \emph{economic constraints} of fraudulent behavior, proposes the \textbf{Fraudster's Trilemma}: organized attackers cannot simultaneously achieve scale, low cost, and dispersed cash-out.
Building on this theory, we derive a robust \emph{structural invariant} in organized fraud---\emph{centralized cash-out}---and use a simple statistical method to turn low-precision individual weak signals into high-precision strong decisions.
The method requires no labels, is nearly parameter-free, white-box interpretable, has linear complexity $O(|E|)$, avoids cold-start issues, and its detection logic possesses the ``open-hand'' property---it can be fully disclosed, because evading the underlying structural invariant forces attackers into economically unattractive operating modes (see Section~\ref{sec:limitations}).

We validate the approach on two real fraud incidents in backtests.
In the promotion abuse case, a single near-zero-cost weak signal (global Precision of only 16\%) after structural amplification achieves Precision above 91\% and unconditional Recall exceeding 99\% ($z{=}10.0$; the signal covers 99.8\% of known fraudsters); at a higher threshold ($z{=}40.0$), Precision reaches 93.7\%.
In the credit card fraud case, an infrastructure-layer weak signal (device spoofing) successfully detects payment-layer attacks without any business-logic linkage---capturing every signal-carrying known fraudster (signal-conditioned Recall 100\%), with unconditional Recall bounded by the signal's 55.9\% coverage; hand-adjudication of flagged-but-unlabeled users estimates true precision $\geq 87\%$ (95\% CI) versus a 17\% nominal lower bound from incomplete police labels---revealing the framework's natural MO-agnostic property, because it relies more on the structural invariant than on signal semantics.
\end{abstract}


% ============================================================
\section{Introduction}\label{sec:intro}

\subsection{Background and Challenges}

Organized platform abuse occurs frequently on large online platforms; typical forms include credit card fraud and promotion abuse.
Such attacks exhibit three salient features: \emph{a large number of involved accounts, high burstiness, and continuously evolving tactics}.
Effectively countering such threats poses a fundamental challenge: existing methods are either insufficiently precise, too costly, or unable to adapt to tactic evolution.

\subsection{Limitations of Existing Approaches}

\textbf{Heuristic rules and single-transaction features.}
Rules or features based on individual transactions are inherently weak signals---when coarse they have low precision; when refined they persistently lag behind updates to attack tactics.
They lack structured cross-account information and yield high false-positive rates.

\textbf{Traditional machine learning.}
Supervised approaches rely on feature engineering and manual annotation, which is costly, and generalize poorly to novel attack tactics.
In fraud settings, reliable labels suffer from \emph{week-to-month-level delays} and are systematically incomplete (see Section~\ref{sec:related-supervised}).

\textbf{Graph models (homogeneous / heterogeneous).}
These incur high engineering complexity and computational cost, depend on seed users, and suffer from severe cold-start problems.

\subsection{Our Approach}

Much risk-control research focuses on model-level improvements---newer architectures (GNN, LSTM, Transformer) and more complex feature engineering.
Yet this path is, in essence, perpetually chasing the attacker's pace.
We take a different strategy: \emph{we start from the economic constraints of fraudulent behavior}.

Rather than asking directly ``what model can detect fraud,'' we first ask: \emph{``What unavoidable constraints does fraudulent behavior face at the economic level?''}
Through an analysis of fraudsters' behavioral incentives, we derive a robust \emph{structural invariant} in organized fraud---regardless of how attack tactics evolve, cost constraints drive fraudsters toward \emph{centralized cash-out} at the fund-exit stage.
This invariant provides a stable, MO-agnostic detection anchor.

On this basis, we design a method of \emph{structural amplification of weak signals}: by exploiting anomalous density accumulation of weak signals at convergence nodes, we convert low-precision individual features into high-precision strong decisions.

\subsection{Main Contributions}

The core contribution of this paper is a \emph{methodological insight}: a correct understanding of the problem structure enables high-precision detection with simple statistical tools.

\begin{enumerate}[leftmargin=*]
  \item \textbf{Fraudster's Trilemma.}
    We identify the fundamental constraint faced by organized attackers---scale, low cost, and dispersed cash-out cannot all be achieved simultaneously---formalizing, for the platform-abuse setting, an asymmetry repeatedly documented in cybercrime-economics measurements (Section~\ref{sec:related-econ}).
    This theory reveals the structural invariant of organized fraud (centralized cash-out) and implies that weak signals tend to accumulate strong statistical concentration at convergence (cash-out) nodes, providing a feasibility basis for simple detection methods.

  \item \textbf{A Framework for structural amplification of weak signals.}
    Building on this structural insight, we design a three-step detection framework (bipartite aggregation $\to$ empirical-Bayes shrinkage $\to$ proportion $Z$-score), achieving high-precision detection at linear complexity.

  \item \textbf{Validation on real cases and cross-modal detection capability.}
    We validate the method on two distinct fraud incidents.
    In Case~1 (promotion abuse), a near-zero-cost weak signal (global Precision of only 16\%) after amplification achieves Precision above 91\% and unconditional Recall exceeding 99\%.
    In Case~2 (credit card fraud) we demonstrate \emph{cross-modal detection capability}---an infrastructure-layer weak signal successfully detects payment-layer attacks without business-logic linkage, revealing the framework's natural MO-agnostic property because it relies more heavily on the structural invariant.
\end{enumerate}

% ============================================================
\section{Related Work}\label{sec:related}

\subsection{Rule-Based and Supervised Learning Methods}\label{sec:related-supervised}

Heuristic rule systems face a precision--flexibility tradeoff between coarse rules and fine-grained rules.
Supervised machine learning faces labeling cost and challenges in generalizing to novel tactics.
Bolton and Hand~\cite{bolton2002statistical} provide a systematic review of statistical fraud detection methods, noting that supervised approaches face fundamental challenges of difficult label acquisition and concept drift.

A critical but often overlooked challenge is the \emph{labeling-delay dilemma}: in fraud detection, reliable labels depend on post-hoc investigations, incurring delays on the order of weeks to months, and labels are incomplete because limited investigation resources can cover only a subset of confirmed cases.

Shared limitations: reliance on single-transaction features, absence of cross-account structural information, and persistent lag behind evolving attacks.

\subsection{Graph-Based Methods}

Pourhabibi et al.~\cite{pourhabibi2020fraud} systematically review applications of graph-based anomaly detection (GBAD) to fraud detection, covering methods based on both static and dynamic graphs.

\textbf{Unsupervised graph methods.}
Beutel et al.~\cite{beutel2013copycatch} (CopyCatch) detect coordinated group attacks on Facebook via temporal lockstep behavior in bipartite graphs.
Jiang et al.~\cite{jiang2014catchsync} (CatchSync) further extend synchronous-behavior detection to large-scale directed graphs.
Hooi et al.~\cite{hooi2016fraudar} (FRAUDAR) propose a camouflage-resistant dense-subgraph detection method and received the KDD 2016 Best Paper Award.
These methods all require global density search or feature decomposition and do not directly support incremental/streaming settings.

\textbf{Dense-block and suspiciousness-metric detectors.}
A closely related line of work formalizes fraud detection as finding dense blocks under a principled suspiciousness metric: CrossSpot~\cite{jiang2015crossspot} scores dense blocks in multimodal behavioral data, M-Zoom~\cite{shin2016mzoom} and D-Cube~\cite{shin2017dcube} accelerate dense-block search in large tensors with approximation guarantees, HoloScope~\cite{liu2017holoscope} combines topological and temporal-spike suspiciousness, and FlowScope~\cite{li2020flowscope} traces multi-step transfer flows to expose money laundering.
The proposed method shares this line's core intuition---fraud concentrates mass in a small substructure---but differs in what is aggregated and why: rather than searching the interaction topology for dense blocks, it aggregates an explicit, cheaply collected weak signal onto the convergence nodes predicted by an economic invariant.
The claimed novelty is accordingly not a new suspiciousness metric, but a deployed, economically anchored, single-signal instantiation whose alert unit and threshold are directly interpretable by fraud-operations teams.

\textbf{Representative GNN work in fraud detection.}
CARE-GNN~\cite{dou2020enhancing} uses reinforcement learning to select neighbors to resist camouflaged fraudsters.
PC-GNN~\cite{liu2021pick} addresses class imbalance via label-balanced sampling.
FraudCenGCL~\cite{lim2025fraudcengcl} employs graph contrastive learning for low-homophily settings.
Shared limitations: engineering complexity, dependence on seeds, and cold-start issues; all require labeled data or seed nodes for training.

\subsection{Economics of Cybercrime}\label{sec:related-econ}

A substantial measurement literature studies cybercrime as a profit-driven value chain and asks which of its stages are scarce and hard to scale.
Levchenko et al.~\cite{levchenko2011click} traced the end-to-end spam value chain and found that the payment tier is its critical chokepoint: the vast majority of spam-advertised purchases cleared through a handful of acquiring banks, which are far harder to replace than domains or hosting.
Follow-up studies of pharmaceutical affiliate programs~\cite{mccoy2012pharmaleaks} and of the commoditization of underground resources~\cite{thomas2015framing} document the same asymmetry: attack-initiation resources (accounts, traffic, infrastructure) are cheap and elastic, whereas monetization channels are scarce, subject to identity and regulatory review, and slow to scale.
The Fraudster's Trilemma (Section~\ref{sec:trilemma}) should be read as a formalization of this recurring empirical pattern for the platform-abuse setting, rather than a wholly new principle: it distills the asymmetry between scalable attack initiation and non-scalable cash-out into an explicit three-way constraint, and derives its structural consequence---centralized cash-out---as a detection anchor.

\subsection{Positioning of the Proposed Method}

The proposed method derives its detection logic from the economic constraints of fraudulent behavior and is independent of the graph neural network paradigm: it requires no complex training process, does not rely on seed nodes, and achieves high-precision detection at linear complexity.
Although it likewise leverages a bipartite graph structure for information aggregation in form (sharing a bipartite-graph perspective with FRAUDAR~\cite{hooi2016fraudar}), the driving force stems from the analysis of attacker behavioral constraints rather than model expressiveness.
The key distinction from unsupervised anomaly detection methods is that the proposed method exploits signal aggregation within graph structure rather than isolated statistical anomalies.

% ============================================================
\section{Methodology}\label{sec:method}

\subsection{Problem Formulation}\label{sec:formulation}

We abstract the trading platform as a \emph{User--Resource Bipartite Graph}:
\begin{equation}
  G = (U \cup V,\; E)
\end{equation}

\begin{itemize}[leftmargin=*]
  \item $U$: set of users / initiating nodes (including normal users and Sybil accounts~\cite{douceur2002sybil})
  \item $V$: set of convergence nodes---broadly, parties on the platform that provide services or goods and receive funds, including merchants, service providers (e.g., ride-hailing drivers), etc.
  \item $E$: transaction edges; $e(u, v)$ denotes a transaction initiated by user $u$ toward node $v$
\end{itemize}

\textbf{Detection objective.}
Without labels and without seed users, identify organized large-scale fraud incidents (organized burst incidents) from massive transaction data.

\textbf{Core challenges.}
Weak signals yield low precision (high recall, low precision); attack tactics evolve continuously; low computational cost is required.

\subsection{Adversarial Economic Model: Behavioral Constraints of Fraudsters}\label{sec:econ-model}

Before delving into the detection algorithm, we first establish a behavioral model of fraudsters.
Unlike ordinary random malicious behavior, organized platform abuse is essentially a \emph{business activity} that pursues a high return on investment (ROI).

\subsubsection{High ROI Necessity}
Organized fraudsters must pursue high ROI---otherwise they could simply run compliant businesses.

\textbf{Cost structure.}
Attackers face multiple costs:
\begin{itemize}[leftmargin=*]
  \item \textbf{One-time fixed costs:} Building technical infrastructure (script development, device farms, proxy IPs), which can be amortized by increasing transaction volume.
  \item \textbf{Per-node recurring costs:} Acquiring cash-out nodes (merchant account registration, bank cards, money mule recruitment); each additional node requires independent investment and cannot be compressed via economies of scale (see Section~\ref{sec:trilemma}).
  \item \textbf{Per-account variable costs:} Acquiring Sybil accounts (purchased accounts, scripted registration) plus labor and time.
\end{itemize}

\textbf{Necessity of scale.}
To cover the above costs and achieve profit, attackers must operate at \emph{scale}---with sufficient Sybil accounts and transaction volume---to amplify total returns.
If sufficient scale cannot be achieved, such abuse is economically infeasible.

\textbf{Knock-on effects of scale.}
Operating at scale requires simultaneously controlling thousands or even tens of thousands of Sybil accounts, which in practice relies on automation (scripts / group control), characteristically leaving \emph{homogeneous behavioral traces}---this is the economic root of the locally high-recall property of weak signals (see Section~\ref{sec:weak-signals}).

\subsubsection{Fast Cash-out Constraint}
Fraudsters are in a race against time with the platform's risk-control system.
Once an attack is exposed, assets will be frozen and all prior investment is lost.
Therefore, attackers must rapidly complete money or value extraction before being identified and banned---i.e., ``fast in, fast out.''

This fast cash-out pressure forces attack behavior to exhibit high \emph{Burstiness} along the time dimension---large volumes of transactions are compressed into short time windows and erupt in concentrated bursts, greatly increasing the difficulty of timely detection and interception.
Because Burstiness and scale naturally co-occur in real attacks (high ROI requires large volume, and time pressure compresses large volume into short windows), the present paper merges the two as the same vertex of the Trilemma.

\subsection{The Fraudster's Trilemma}\label{sec:trilemma}

Based on the above economic logic, we propose the \textbf{Fraudster's Trilemma}---a formalization, for the platform-abuse setting, of the scalable-initiation versus non-scalable-monetization asymmetry repeatedly documented in the cybercrime-economics literature (Section~\ref{sec:related-econ}).
For any organized attack group, the following three objectives \emph{cannot be achieved simultaneously}:

\begin{enumerate}[leftmargin=*]
  \item \textbf{Scalability \& Burstiness}---the fraud operation is scalable overall---able to amplify returns by increasing Sybil accounts and transaction volume (the economic need for scale)---and erupts in concentrated bursts within short time windows to complete cash-out before risk control intervenes (Burstiness).
  \item \textbf{Low Cost}---controlling the total operating cost of the attack---including Sybil account acquisition, technical infrastructure (device farms, proxy IPs), labor, and acquisition and maintenance of cash-out nodes.
  \item \textbf{Cash-out Dispersion}---dispersing fund exits across many independent cash-out nodes (merchant accounts, bank cards, money mule accounts) to evade detection of convergence patterns.
\end{enumerate}

\textbf{Root of the constraint---asymmetry between Scalability and economies of scale.}
Different stages of a fraud operation behave very differently in terms of scalability.
The attack initiation side (Sybil accounts, scripts, device farms) is naturally horizontally scalable and enjoys positive economies of scale---marginal cost decreases with scale; however, neither class of cash-out node on the fund-exit side possesses this scalability:

\begin{itemize}[leftmargin=*]
  \item \textbf{Collusive nodes} (merchant accounts, bank cards, money mules): each requires independent KYC review or real-name registration; each mule requires independent recruitment and maintenance; and qualified identities are themselves scarce resources---unit acquisition cost does not fall with scale and may even rise with demand.
  \item \textbf{Resale nodes} (merchants of specific goods): the number of merchants supplying high-value, easily liquidated goods (e-gift cards, consumer electronics) is naturally limited; large-scale centralized purchasing easily triggers merchant-side risk controls; and secondary-market resale channels also have capacity ceilings.
\end{itemize}

The common characteristic of both classes of cash-out nodes is: \emph{not horizontally scalable, no positive economies of scale, and possibly even diseconomies of scale.}
Under the economic necessity of operating at scale, maintaining Cash-out Dispersion requires a proportional increase in cash-out nodes, and this portion of cost cannot be compressed via economies of scale---\emph{therefore, under scaled operations, Cash-out Dispersion and Low Cost are in direct tension.}

\textbf{Scope conditions.}
The load-bearing assumption above is that \emph{per-node cash-out cost does not decrease with scale}.
We state its boundaries explicitly.
The assumption is strongest where cash-out must pass platform- or bank-side identity review---KYC-verified merchant accounts, bank cards, mule recruitment---which covers the settings studied in this paper.
It may weaken where monetization can partially escape such review: cryptocurrency off-ramps with lax controls, or highly automated resale of digital goods, can restore some economies of scale on the exit side.
In such settings the Trilemma's tension is diluted, the many-to-few topology may be less pronounced, and the method's effectiveness is expected to degrade accordingly (consistent with the limitation stated in Section~\ref{sec:limitations}).

\begin{figure}[t]
  \centering
  \includegraphics[width=\columnwidth]{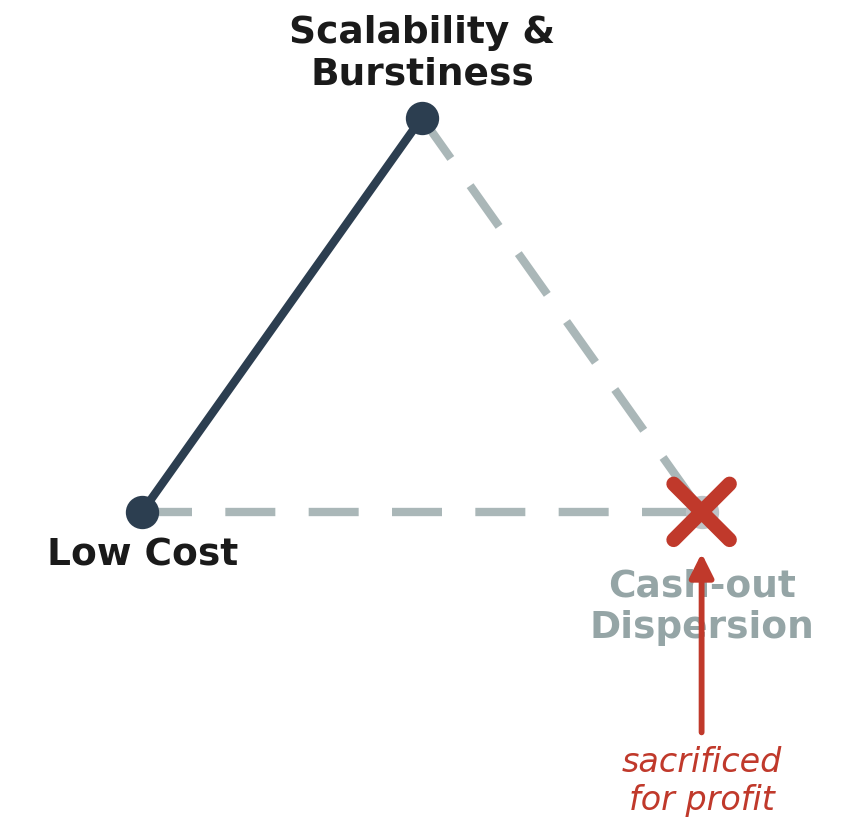}
  \caption{The Fraudster's Trilemma. Rational attackers pursuing profit must sacrifice cash-out dispersion (crossed out) due to asymmetric scalability: the Sybil supply side enjoys economies of scale, while cash-out nodes do not.}
  \label{fig:trilemma}
\end{figure}

\subsection{Structural Invariant: Centralized Cash-Out}\label{sec:invariant}

A direct corollary of the Trilemma: rational attackers pursuing high ROI (high scale + low cost) are forced to \emph{reuse a small number of cash-out nodes}---sacrificing ``Cash-out Dispersion.''
We name this corollary \textbf{centralized cash-out} and regard it as a robust \emph{structural invariant} in organized fraud.
Regardless of how attack tactics evolve, economic pressure preserves this \emph{many-to-few} convergence topology (its detection implications are detailed in Section~\ref{sec:amplification}).

\begin{figure}[t]
  \centering
  \includegraphics[width=\columnwidth]{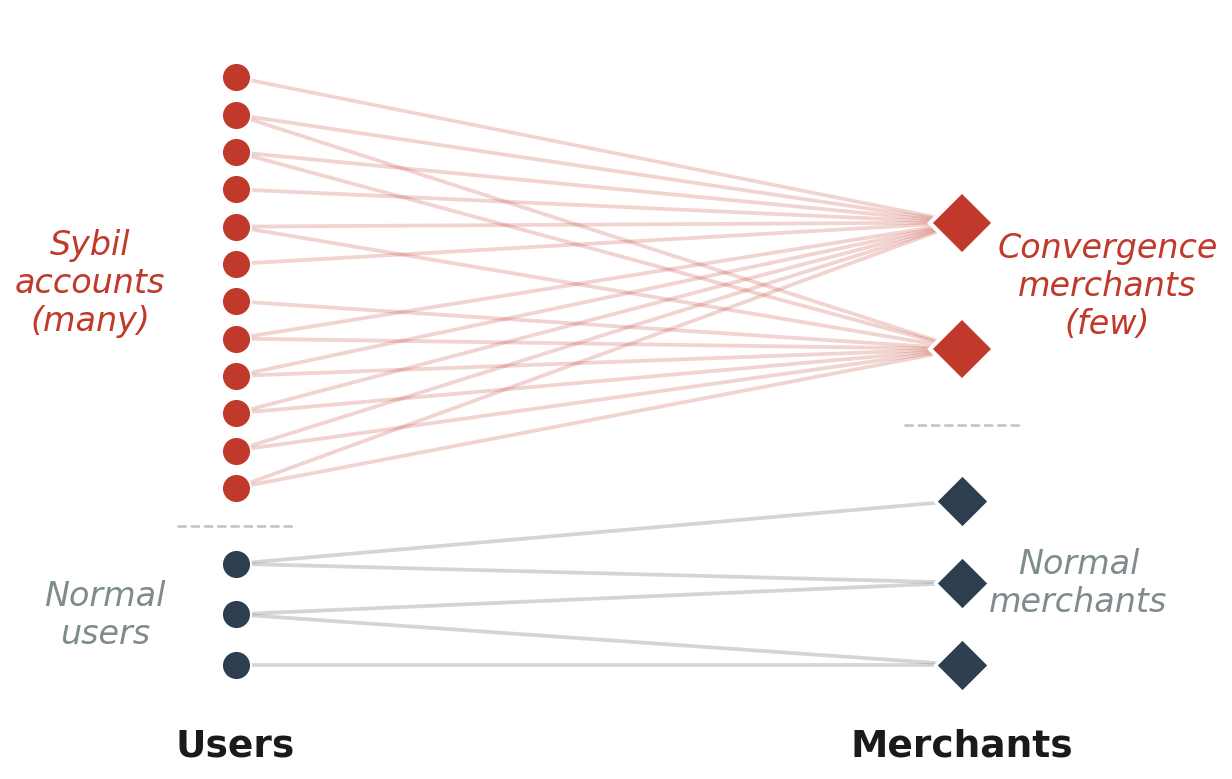}
  \caption{Bipartite convergence structure. Many Sybil accounts (left, red) concentrate transactions on few convergence merchants (right, red diamonds), forming the many-to-few topology predicted by the Trilemma. Normal users connect to normal merchants with no cross-contamination.}
  \label{fig:bipartite}
\end{figure}

\subsection{Definition and Characteristics of Weak Signals}\label{sec:weak-signals}

\subsubsection{Definition}
\textbf{Weak signals} are weak anomalous features on the transaction initiation side (Sybil/User side) that can be \emph{collected at low cost} based on empirical rules.

A weak signal must first be an anomalous signal; its ``weakness'' lies in the fact that, when used alone, precision is insufficient for direct adjudication.

Typical examples: device in a jailbroken state; multiple accounts logged in on the same device (Device Sharing); use of a Promotion Code to cover most (or even all) of the transaction cost; IP address belonging to a non-residential segment or use of proxy/VPN.

\subsubsection{Binary Paradox}
Weak signals exhibit a \emph{binary paradox} (Table~\ref{tab:binary-paradox}).

\begin{table}[t]
  \centering
  \caption{The binary paradox of weak signals.}
  \label{tab:binary-paradox}
  \small
  \begin{tabular}{@{}p{1.8cm}p{3.4cm}p{2cm}@{}}
    \toprule
    \textbf{Property} & \textbf{Root cause} & \textbf{Implication} \\
    \midrule
    Local High Recall & \textbf{Inevitable cost of scaled profit-seeking} (Section~\ref{sec:econ-model}): the attacker must control thousands to tens of thousands of Sybil accounts; batch operations rely on scripts/group-control software; scale-driven automation leads to strong homogeneity of weak signals within the same attack batch & Extremely high coverage within attack samples \\
    \addlinespace
    Global Low Precision & Similar features also appear at low probability in the normal user population (power users with jailbroken phones, family members sharing devices, etc.); blocking directly on this basis would cause a high false-positive rate & Insufficient alone as a basis for account suspension \\
    \bottomrule
  \end{tabular}
\end{table}

\textbf{Core tension:} Weak signals are highly consistent within attacks (high recall), but are too noisy from a global perspective (low precision), and therefore cannot be used alone for adjudication.

\subsection{Core Mechanism: Structural Amplification of Weak Signals}\label{sec:amplification}

Funds or value must eventually leave the system---the cash-out convergence node is a bottleneck the attacker cannot bypass.
Combining the locally high-recall property of weak signals in Section~\ref{sec:weak-signals} with the centralized cash-out structure (many-to-few topology) in Section~\ref{sec:invariant}, we state the core detection hypothesis of this paper:

\subsubsection{Anomaly Concentration at Cash-out Nodes}

\begin{quote}
When large-scale organized fraud occurs, a large number of Sybil accounts carrying the same anomalous signals (weak signals) must connect to a small number of cash-out nodes, causing \emph{statistically implausible density accumulation} of anomalous signals at those nodes.
\end{quote}

\textbf{Intuition behind the weak-to-strong transition:}
\begin{itemize}[leftmargin=*]
  \item A single account on a jailbroken device $\to$ may be coincidence (weak signal)
  \item Among 1{,}000 receipt transactions to one merchant, 99\% originate from jailbroken devices $\to$ this is no longer statistical fluctuation, but \emph{near-certain attack adjudication (strong signal)}
\end{itemize}

\subsubsection{Mathematical Formalization}

This detection logic essentially exploits the \emph{Law of Large Numbers} (LLN) and the \emph{Central Limit Theorem} (CLT): LLN ensures that, for normal merchants, the proportion of weak signals stabilizes toward a global baseline rate as transaction volume grows; CLT motivates a volume-normalized $Z$-score that quantifies the deviation (see Section~\ref{sec:algorithm}, Step~3).

Let the natural occurrence probability of a weak signal $s$ in the normal user population be $p$ ($0 < p \ll 1$, small but nonzero).
\begin{itemize}[leftmargin=*]
  \item \textbf{Normal merchant:} among $n$ associated users, the expected number of occurrences of signal $s$ is $\approx np$ (background noise level).
  \item \textbf{Anomalous convergence node:} because funding sources are dominated by homogeneous fraudulent accounts, the signal frequency deviates sharply from the expectation and approaches $n$.
\end{itemize}

\textbf{Key observation:} when any weak signal occurs naturally as background noise, its occurrence rate should be fixed---\emph{within a unit of time, the number of weak signal occurrences at a merchant should be proportional to total transaction volume}.
When, for some merchant, the count of a given weak signal is markedly larger than the expectation implied by total transaction volume, it is readily identifiable as anomalous by scoring the deviation from that expectation.

Through this \emph{structural aggregation}, by measuring the deviation of weak signal density at convergence nodes from the expected level, massive low-confidence individual features can be converted into high-confidence structured risk scores.

\begin{figure}[t]
  \centering
  \includegraphics[width=\columnwidth]{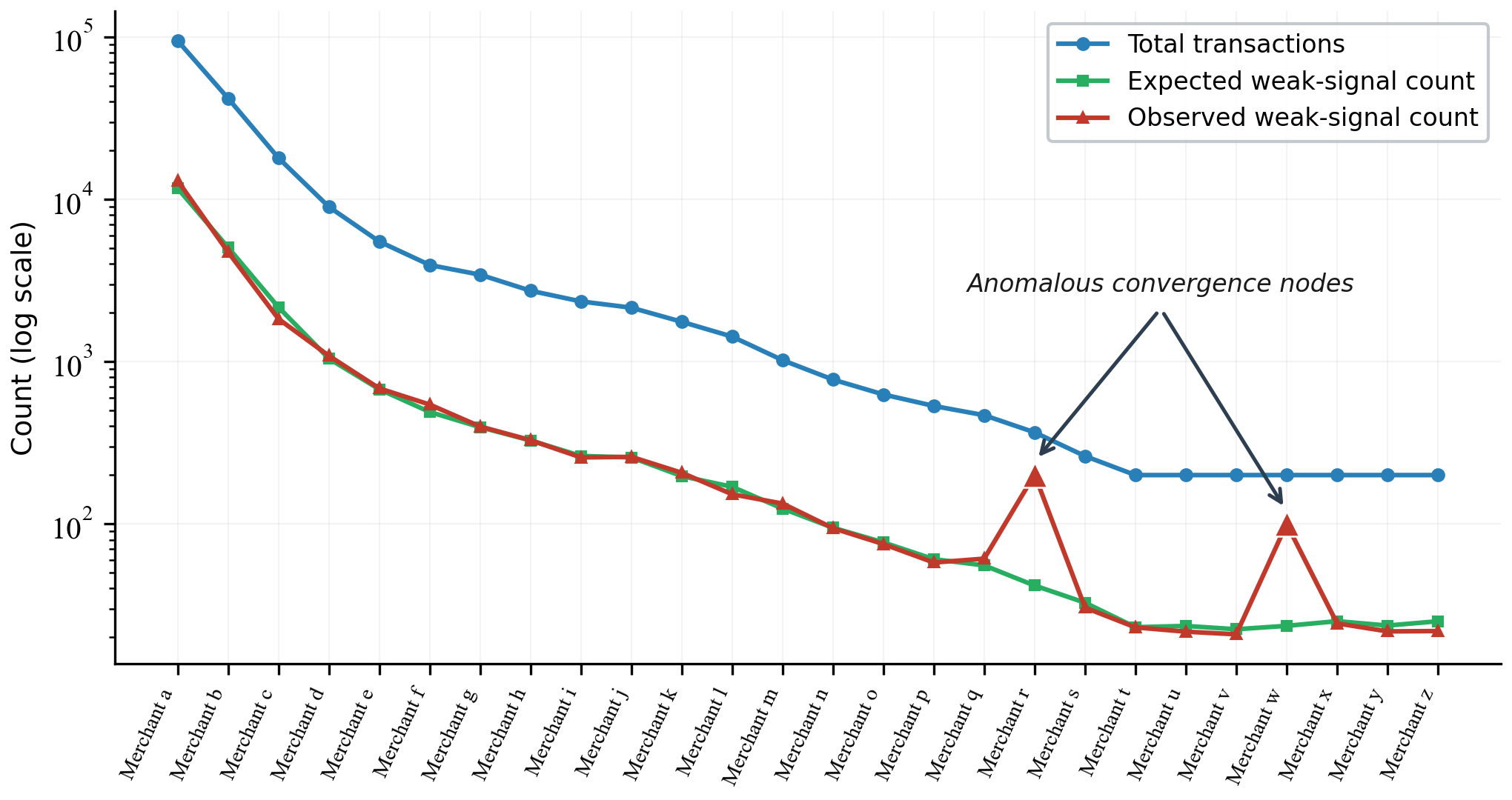}
  \caption{Structural amplification mechanism (illustrative). Merchants sorted by total transaction volume (blue). The expected weak-signal count (green) tracks total volume proportionally. At two anomalous merchants (Merchant~r, Merchant~w), the actual observed count (red) spikes far above the expected baseline---these are the convergence nodes surfaced by the $Z$-score.}
  \label{fig:longtail}
\end{figure}

\subsection{Algorithm Design}\label{sec:algorithm}

Based on the preceding theoretical derivation, the concrete detection algorithm requires only three steps.
The contribution of this paper lies at the methodological level---once the structure of the problem is correctly understood, simple statistical tools suffice.

\subsubsection{Step 1: Bipartite Aggregation}
For each weak signal $f$, aggregate the user-side binary label along transaction edges to the convergence node $v$:
\begin{equation}
  s_v = \sum_{u \in \mathcal{N}(v)} f(u), \quad t_v = |\mathcal{N}(v)|
\end{equation}
where $s_v$ is the number of weak signal hits in transactions associated with node $v$, and $t_v$ is the total number of transactions.

The framework is designed to be \emph{entity-agnostic}: $v$ may be a merchant, a driver, or any convergence node role, and the signal $f$ can be plug-and-play replaced.

\subsubsection{Step 2: Empirical Bayes Smoothing}
The na\"ive proportion estimate $\hat{p}_v = s_v / t_v$ is highly unstable for small-sample nodes (a node with only two transactions may coincidentally hit the signal on both, yielding a raw ratio of 100\%).
We introduce a shrinkage estimate:
\begin{equation}\label{eq:shrinkage}
  \tilde{p}_v = \frac{s_v + M \cdot p_{\mathrm{global}}}{t_v + M}
\end{equation}

\begin{itemize}[leftmargin=*]
  \item $p_{\mathrm{global}}$: global weak signal occurrence rate (prior baseline)
  \item $M = \bar{t}$: global mean transaction volume, serving as adaptive prior strength
\end{itemize}

Intuition: this is equivalent to injecting $M$ ``virtual observations'' for each node, with observed value equal to the global baseline rate.
Nodes with large transaction volume ($t_v \gg M$) are almost unaffected (data dominate); nodes with small transaction volume are pulled back toward the prior (noise suppression).
This is a standard application of the James--Stein estimator in the binomial-proportion setting~\cite{efron2010large}.\footnote{Formally, Eq.~\eqref{eq:shrinkage} is the exact posterior mean of $p_v$ under a $\mathrm{Beta}\!\left(M p_{\mathrm{global}},\, M(1-p_{\mathrm{global}})\right)$ prior (prior mean $p_{\mathrm{global}}$, prior strength $M$) with a binomial likelihood; the smoothing is thus an empirical-Bayes posterior mean rather than an ad hoc device.}

\subsubsection{Step 3: Proportion \texorpdfstring{$Z$}{Z}-Score}

\textbf{Motivation for the score's form:} under an idealized null model in which each transaction's weak signal hit is an independent Bernoulli trial (success probability $p_{\mathrm{global}}$), the total number of hits $s_v$ among $t_v$ transactions at node $v$ follows a binomial distribution $B(t_v, p_{\mathrm{global}})$, and the standardized deviation of the observed proportion from the background rate takes the familiar one-proportion form:

\begin{equation}\label{eq:ztest}
  z_v = \frac{\tilde{p}_v - p_{\mathrm{global}}}{\sqrt{p_{\mathrm{global}}(1 - p_{\mathrm{global}}) \;/\; t_v}}
\end{equation}

\textbf{Interpretation as an anomaly score, not a calibrated test statistic:} the independence assumption behind the binomial null does not hold in the very regime this paper targets---at a fraudulent convergence node, signal hits are strongly positively correlated (shared device farms, scripted behavior, homogeneous account profiles), which is precisely the premise of Section~\ref{sec:weak-signals}.
Correlation shrinks the effective sample size below $t_v$ and inflates $z_v$ relative to its nominal null distribution, which is why incident-period values can reach the hundreds (Table~\ref{tab:case2-signals}).
We therefore do \emph{not} interpret $z_v$ through normal-tail probabilities or $p$-values.
Instead, $z_v$ serves as an \emph{anomaly ranking score}: it is monotone in the deviation of a node's signal density from the global background, normalized by transaction volume, and the operating threshold is chosen empirically above the calm-period noise floor (Section~\ref{sec:method-exp}) rather than from normal quantiles.

A large positive $z_v$ indicates that the concentration of the weak signal at node $v$ far exceeds the global background---that is, the node is a high-risk convergence node.

\emph{Note:} the numerator uses the shrinkage estimate $\tilde{p}_v$ while the denominator keeps the un-shrunk null standard error.
This combination is \emph{uniformly conservative}: since $\tilde{p}_v - p_{\mathrm{global}} = \frac{t_v}{t_v + M}\,(\hat{p}_v - p_{\mathrm{global}})$, the score equals the un-shrunk score scaled by $\frac{t_v}{t_v+M} \le 1$.
Shrinkage can therefore only lower a node's score, never raise it, and the penalty is concentrated exactly on small-sample nodes---preventing them from producing false positives due to noise.

\subsubsection{Signal-Agnostic Composability}

The choice of weak signals is open: each weak signal independently executes the three steps above, and detection outputs are merged after being produced independently.
Failure of a single signal does not undermine overall detection capability.

\textbf{Attack-vector agnosticism:} This composable design frees the framework from needing to know in advance which weak signal the current attack relies on.
Structural aggregation itself acts as an unsupervised feature selector---when a class of attack occurs, the associated weak signals automatically produce pronounced density deviations at convergence nodes and are thus ``lit up.''
The system does not require manual specification of ``which signal to focus on for this attack''; the aggregation phenomenon completes the selection automatically.

\subsubsection{Algorithmic Properties}

\begin{table}[t]
  \centering
  \caption{Algorithmic properties of the proposed method.}
  \label{tab:properties}
  \small
  \begin{tabular}{@{}p{2.8cm}p{4.8cm}@{}}
    \toprule
    \textbf{Property} & \textbf{Description} \\
    \midrule
    Approximately nonparametric & $M$ is determined adaptively from data; the only choice required is the $z$ threshold \\
    \addlinespace
    Linear complexity $O(|E|)$ & Each step is a single pass over transaction edges \\
    \addlinespace
    Training-free & No model fitting; ready to use out of the box \\
    \addlinespace
    Entity-agnostic & The same framework applies to different types of convergence nodes \\
    \addlinespace
    Incremental update $O(1)$ & Each new transaction only needs to update $s_v$ and $t_v$ for the corresponding node and recompute $z_v$; naturally suited to streaming \\
    \addlinespace
    Low-cost signal extension & Adding a weak signal only requires defining a binary function $f(u)$; no retraining and no framework modification needed \\
    \bottomrule
  \end{tabular}
\end{table}

\textbf{Real-time deployment feasibility:} all three computational steps of the algorithm can be implemented as incremental operations---when each transaction arrives, only two counter updates and one $Z$-score recomputation are needed for the target convergence node, with $O(1)$ per-update complexity, naturally suited to a streaming architecture.
Experiments in this paper use daily batch turn-based replay (Daily Batch); see Section~\ref{sec:method-exp}.

% ============================================================
\section{Experiments}\label{sec:experiments}

\subsection{Experimental Methodology}\label{sec:method-exp}

Because timely intervention by the detection system prevents attacks from continuing to scale, online A/B experiments face a fundamental limitation: one cannot deliberately allow known attacks through to collect control-group data.
Therefore, this paper adopts two complementary evaluation approaches:

\begin{enumerate}[leftmargin=*]
  \item \textbf{Backtesting:} We select historically occurred fraud incidents that were recorded ex post, replay transaction data in chronological order, and simulate a real-time detection setting.
  \item \textbf{Live Deployment:} In Case~1, the system was connected during the incident and went live, providing detection performance under a genuine online environment.
\end{enumerate}

\textbf{Backtesting granularity:} Turn-based daily batches; every day, $Z$-scores are recomputed for all convergence nodes.

\textbf{Specificity and the calm-period noise floor:} During attack-free calm periods, $z$-scores are not identically zero---residual correlation in normal traffic keeps the maximum calm-period $z$ at a stable \emph{noise floor} (approximately 20--30 in Case~2; Table~\ref{tab:case2-signals}).
Empirically, across calm-period backtests covering more than 1{,}000 convergence nodes over more than 30 days (over $3{\times}10^4$ node-days), the score never exceeded ${\sim}30$.
Because the operating threshold is referenced to this \emph{maximum over all nodes and days}, the multiplicity of scanning many nodes daily is already absorbed into the empirical noise floor, and no separate multiple-testing correction is applied.
The recommended operating threshold $z{=}40$ is chosen \emph{above} this noise floor: at that operating point, the number of users flagged per day approaches zero during calm periods, and incident-period scores (in the hundreds) separate from the floor by nearly an order of magnitude.
The daily time series of flagged users in each case (Fig.~\ref{fig:case1-daily}, Fig.~\ref{fig:case2-atu}b) provides intuitive confirmation: comparing before and after the attack, no alerts are produced during calm periods at the operating threshold.

\subsection{Case~1: Market~A Promo Abuse Incident}\label{sec:case1}

\subsubsection{Incident Background}
In the second half of 2025, Southeast Asia Market~A experienced an organized promo-abuse incident.
Attackers registered Sybil accounts in bulk and abused Promo Codes, using coupons to cover all or most of the trip cost (i.e., ``something for nothing''), then completed centralized cash-out on the driver side through fictitious trips with collusive drivers.

This is a typical \emph{non-injection fraud}---no external capital is injected into the system, and Sybil account balances are effectively zero or near zero.

\textbf{Attack scale:}
\begin{itemize}[leftmargin=*]
  \item Sybil accounts involved: thousands (confirmed blocked after manual review)
  \item Drivers involved (Cash-out side): low hundreds (cumulatively confirmed fraudulent drivers); the system cumulatively flagged ($z{\geq}40.0$) on the order of hundreds
  \item Attack duration: several months
\end{itemize}

\textbf{Limitations of traditional detection:}
In such promo-abuse incidents, existing risk controls mainly rely on trip-level anomaly detection---such as GPS trajectory forgery identification, route plausibility checks, and abnormal trip-duration rules.
These methods are essentially per-transaction feature rules; when attackers use real trips (actual travel with collusive drivers) or continuously adjust forgery strategies, both precision and coverage degrade rapidly.
The proposed method bypasses tactic-level confrontation at the transaction layer and instead operates directly on the fund-convergence structure.

\subsubsection{Amplification of a Near-Zero-Cost Weak Signal}

The core weak signal in this case is \texttt{use\_promo} (whether a transaction used promotional subsidy), an almost \emph{zero-collection-cost} binary feature---extracted directly from existing fields in transaction records, requiring no additional data sources or model inference.

Using this signal alone as a blocking rule would cause catastrophic collateral damage---normal users also enjoy high subsidy ratios during major platform promotions.
However, after structural amplification, this signal exhibits extremely strong discriminative power at convergence nodes.

\textbf{No-amplification baseline (Ablation):}
In a single-day snapshot, tens of thousands of users carry this signal, of whom thousands are confirmed fraudsters.
If signal hits are used directly as a blocking rule (without structural amplification), global Precision is only about 16\%.
After structural amplification, Precision increases to 93\%+, an \emph{amplification factor of approximately $6{\times}$}.

\textbf{Multi-signal comparison:}
In the same incident, two weak signals both achieve high Precision after amplification, but their signal coverage differs substantially (Table~\ref{tab:case1-signals}).

\begin{table}[t]
  \centering
  \caption{Case~1: Multi-signal comparison at $z{=}10.0$.}
  \label{tab:case1-signals}
  \small
  \begin{tabular}{@{}lcccc@{}}
    \toprule
    \textbf{Weak signal} & \textbf{Coverage} & \textbf{Fraudsters} & \textbf{Precision} & \textbf{SCR} \\
    \midrule
    \texttt{use\_promo} & \textbf{99.82\%} & 3331/3337 & 91.01\% & 99.73\% \\
    \texttt{device\_spoofing} & 25.14\% & 839/3337 & 92.83\% & $\approx$98.8\% \\
    \bottomrule
  \end{tabular}
\end{table}

After structural amplification, the two signals achieve \emph{comparable Precision and SCR} (Precision ${>}91\%$, SCR ${>}98\%$), demonstrating that the amplification mechanism is equally effective across different signals.
The key difference lies in \emph{signal coverage}: \texttt{use\_promo} covers 99.82\% of fraudsters, whereas \texttt{device\_spoofing} covers only 25.14\%.
This directly reflects the MO-agnostic framework's ``selective activation'': for promo abuse, the business-layer signal naturally becomes the highest-coverage detector; while the device-layer signal is amplified equally strongly, its coverage is limited because it bears no direct relation to this attack's tactics---conversely, in the forthcoming Case~2 (credit-card fraud), \texttt{device\_spoofing} becomes the dominant signal.

\subsubsection{Quantitative Evaluation}\label{sec:case1-quant}

\textbf{Ground-truth definition:} True positives are defined as accounts \emph{blocked after manual review} (thousands in total); labeling delay is relatively small, making Precision estimates more reliable.

\textbf{Single-day snapshot detection performance} (single cross-section during the attack peak; the denominator of Signal-conditioned Recall is the number of fraudsters carrying the signal, 3{,}331):

\begin{table}[t]
  \centering
  \caption{Case~1: Single-day snapshot (use\_promo).}
  \label{tab:case1-pr}
  \small
  \begin{tabular}{@{}cccccc@{}}
    \toprule
    $z$ & \textbf{HR Drivers} & \textbf{Flagged Users} & \textbf{Caught} & \textbf{Prec.} & \textbf{SCR} \\
    \midrule
    1.0  & 157 & 3956 & 3329 & 84.15\% & 99.94\% \\
    5.0  & 120 & 3843 & 3329 & 86.63\% & 99.94\% \\
    10.0 &  84 & 3650 & 3322 & 91.01\% & 99.73\% \\
    \textbf{40.0} & \textbf{59} & \textbf{3196} & \textbf{2994} & \textbf{93.68\%} & \textbf{89.88\%} \\
    \bottomrule
  \end{tabular}
\end{table}

Signal coverage $= 3{,}331 / 3{,}337 = 99.82\%$ (almost all known fraudsters carry this signal).
Unconditional Recall $=$ SCR $\times$ signal coverage; hence at $z{=}10.0$, Unconditional Recall $= 99.73\% \times 99.82\% \approx 99.55\%$.

\textbf{Key observations:}

\begin{enumerate}[leftmargin=*]
  \item \textbf{At threshold $z{=}10.0$, Precision already exceeds 91\%; at $z{=}40.0$, it reaches 93.68\%}---fully validating the effectiveness of weak signal structural amplification. A standalone binary feature with virtually no discriminative power (global Precision of only 16\%) achieves high-precision detection after bipartite aggregation, an amplification of approximately $6{\times}$.

  \item \textbf{Signal-conditioned Recall reaches 99.94\% at $z{\leq}5.0$} (3{,}329/3{,}331), validating the \emph{locally high-recall} property of weak signals. At $z{=}40.0$, SCR drops to 89.88\%, reflecting the Precision--Recall tradeoff when the attack is distributed across many medium-risk drivers. Because signal coverage is as high as 99.82\% (3{,}331/3{,}337), Unconditional Recall is nearly equal to SCR.

  \item \textbf{The Precision--SCR tradeoff is clear and controllable:} from $z{=}10.0$ (P=91\%, SCR=99.7\%) to $z{=}40.0$ (P=93.7\%, SCR=89.9\%), raising the threshold trades approximately 10\% SCR for approximately 3\% Precision gain, allowing operations teams to flexibly choose their operating point.

  \item \textbf{Operational feasibility:} The system uses convergence nodes as the alert unit; each alert is accompanied by an associated list of suspicious users. Under the recommended threshold $z{=}40.0$, no alerts are produced during calm periods (specificity verification in Section~\ref{sec:method-exp}); during the incident, the number of alerted convergence nodes per day is at most on the order of tens (59 driver nodes on the sampled day in Case~1; only 7 merchant nodes on the sampled day in Case~2). Although associated users can number in the thousands, analysts' review workload is determined by the number of nodes rather than the number of users---because Precision is extremely high (93.68\%), nearly every alert is a true positive, making the review workflow efficient without requiring extensive false-positive triage. By contrast, if weak signal hits were used directly as alert triggers (without structural amplification), more than 20{,}000 user-level alerts would be produced in a single day with Precision of only 16\%, making manual review infeasible in practice.
\end{enumerate}

\subsubsection{Live Deployment and Temporal Validation}

\begin{figure}[t]
  \centering
  \includegraphics[width=\columnwidth]{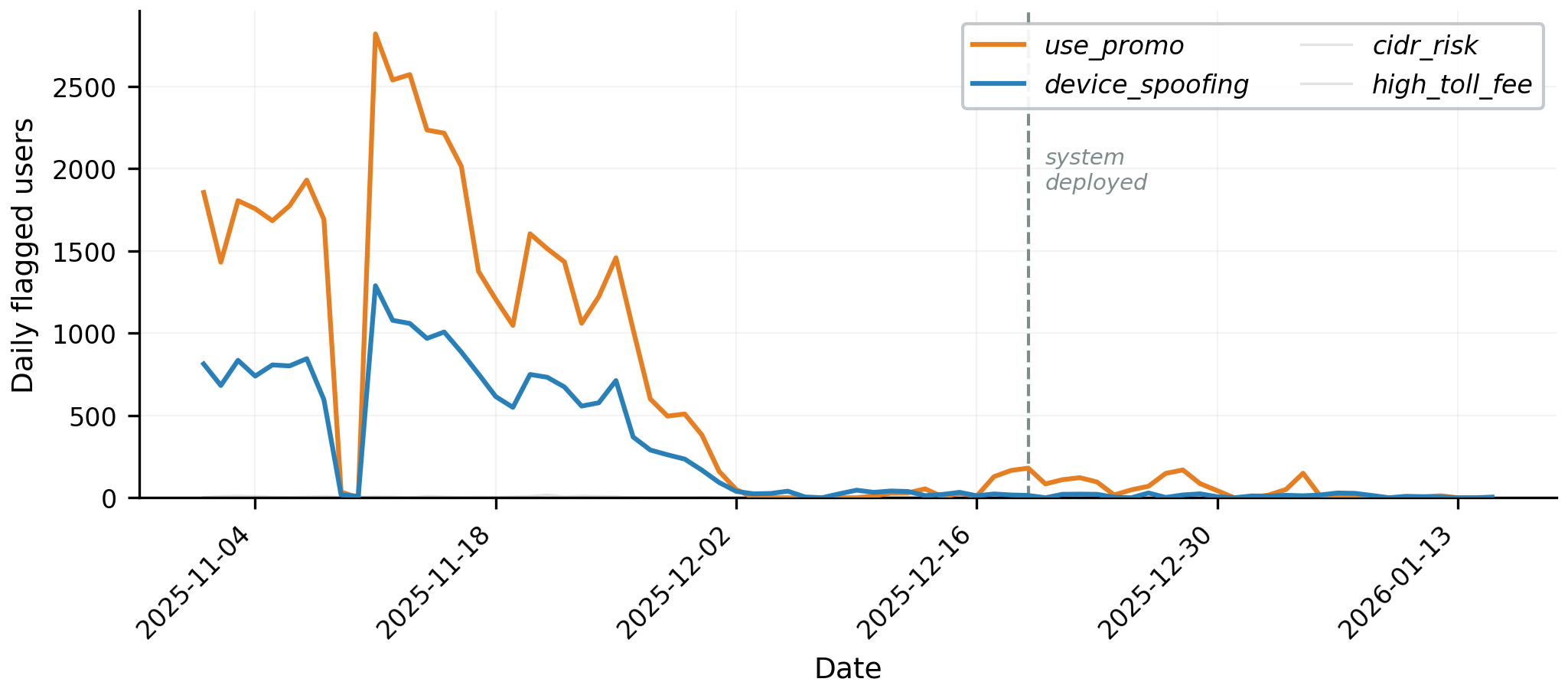}
  \caption{Case~1: Daily flagged users by weak signal ($z{\geq}40.0$). The decline in late November reflects business-side interventions (promo paused), not system deployment. The system went live on December 19, 2025 (dashed line). Post-deployment, the system detected residual attackers (peaking ${\sim}300$/day in late December and early January, then sporadic detections), enabling timely enforcement blocking; the attack never recovered to its October--November peak. \texttt{use\_promo} (orange) consistently flags more users than \texttt{device\_spoofing} (blue), matching the single-day coverage gap in Section~\ref{sec:case1-quant}. Other signals (\texttt{cidr\_risk}, \texttt{high\_toll\_fee}, gray) remain inactive throughout---evidence of selective activation.}
  \label{fig:case1-daily}
\end{figure}

Fig.~\ref{fig:case1-daily} shows the daily number of flagged users ($z \geq 40$) for multiple weak signals during the incident.

\textbf{Event timeline:}
\begin{itemize}[leftmargin=*]
  \item \textbf{October to early-mid November:} Attack at peak; backtesting shows daily flagged passengers in the thousands
  \item \textbf{Late November:} Risk team suppressed the attack through business-side interventions (pausing specific promo campaigns), daily flagged users began to decline
  \item \textbf{December:} Promo campaigns gradually resumed, with risk of attack resurgence
  \item \textbf{December 19:} The system went live, connected to the real-time risk workflow (concurrent with other parallel rules by the risk team to suppress the attack)
  \item \textbf{Late December to January:} The system detected residual attackers (peaking ${\sim}300$/day in late December and early January, then sporadic detections), enabling the operations team to block them promptly; the attack never recovered to its October--November peak
\end{itemize}

\textbf{Key observations:}

\begin{enumerate}[leftmargin=*]
  \item \textbf{Signal coverage differences remain stable over time:} \texttt{use\_promo} (orange) consistently flags substantially more users per day than \texttt{device\_spoofing} (blue) throughout the incident, consistent with the signal coverage gap in the single-day snapshot in Section~\ref{sec:case1-quant} (99.82\% vs 25.14\%). This indicates that the coverage difference is not a single-day sampling artifact but a stable characteristic throughout the entire incident.

  \item \textbf{Timely detection and response after deployment:} After the system went live on December 19, the system detected residual attackers (peaking ${\sim}300$/day in late December and early January, then sporadic detections). Combined with prompt enforcement blocking by the operations team, this validates the system's operational feasibility in a real production environment.

  \item \textbf{Longitudinal validation of selective activation:} Other weak signals shown in the same figure (e.g., \texttt{cidr\_risk}, \texttt{high\_toll\_fee}) produce almost no detections throughout the incident, further validating the framework's MO-agnostic property---only signals related to the current attack tactics are ``activated.''
\end{enumerate}

\subsection{Case~2: Market~B Fraudulent Top-Up Incident}\label{sec:case2}

\subsubsection{Incident Background}
During Q1--Q2 2025, Southeast Asia Market~B experienced an organized credit-card fraud incident.
Attackers exploited design weaknesses in the platform's top-up mechanism, bypassed user-side verification, bound stolen credit cards to Sybil accounts to perform fraudulent top-ups, and then completed centralized cash-out through fictitious transactions with collusive merchants.

\textbf{Attack scale:}
\begin{itemize}[leftmargin=*]
  \item Sybil accounts involved: low hundreds confirmed on the police report list; the system cumulatively flagged on the order of thousands
  \item Collusive merchants involved (Cash-out side): fewer than ten (approximately 7 merchants with $z{\geq}40.0$ in the single-day snapshot)
  \item Attack duration: approximately two months
\end{itemize}

\textbf{Limitations of traditional detection:}
During the incident, the risk team's emergency response was to write dedicated rules targeting fraudulent top-up behavior---setting thresholds based on payment-layer features such as top-up counts, failure ratios, and per-transaction amounts.
However, such rules inherently face a precision--recall dilemma, and attackers can adjust top-up patterns to evade specific thresholds, resulting in sustained adversarial attrition.
The proposed method bypasses payment-layer tactic confrontation and detects the convergence structure dictated by economic constraints (see Section~\ref{sec:limitations}).
As shown in Section~\ref{sec:cross-modal}, even an infrastructure-layer signal with zero connection to payment logic can capture the attack.

\subsubsection{Cross-Modal Detection}\label{sec:cross-modal}

The most striking finding in this case is that \emph{an infrastructure-layer weak signal successfully detected a business-layer attack}.

The best-performing weak signal in backtesting is \texttt{device\_spoofing} (device-environment anomaly determination, encompassing generic features such as jailbroken devices, emulators, and multi-account device sharing)---a long-standing, generic infrastructure-layer signal with no direct business-logic connection to credit-card fraud.
However, because attackers reused underlying device infrastructure to control costs, this generic signal formed a pronounced statistical concentration at collusive merchant nodes.

\begin{figure}[t]
  \centering
  \includegraphics[width=\columnwidth]{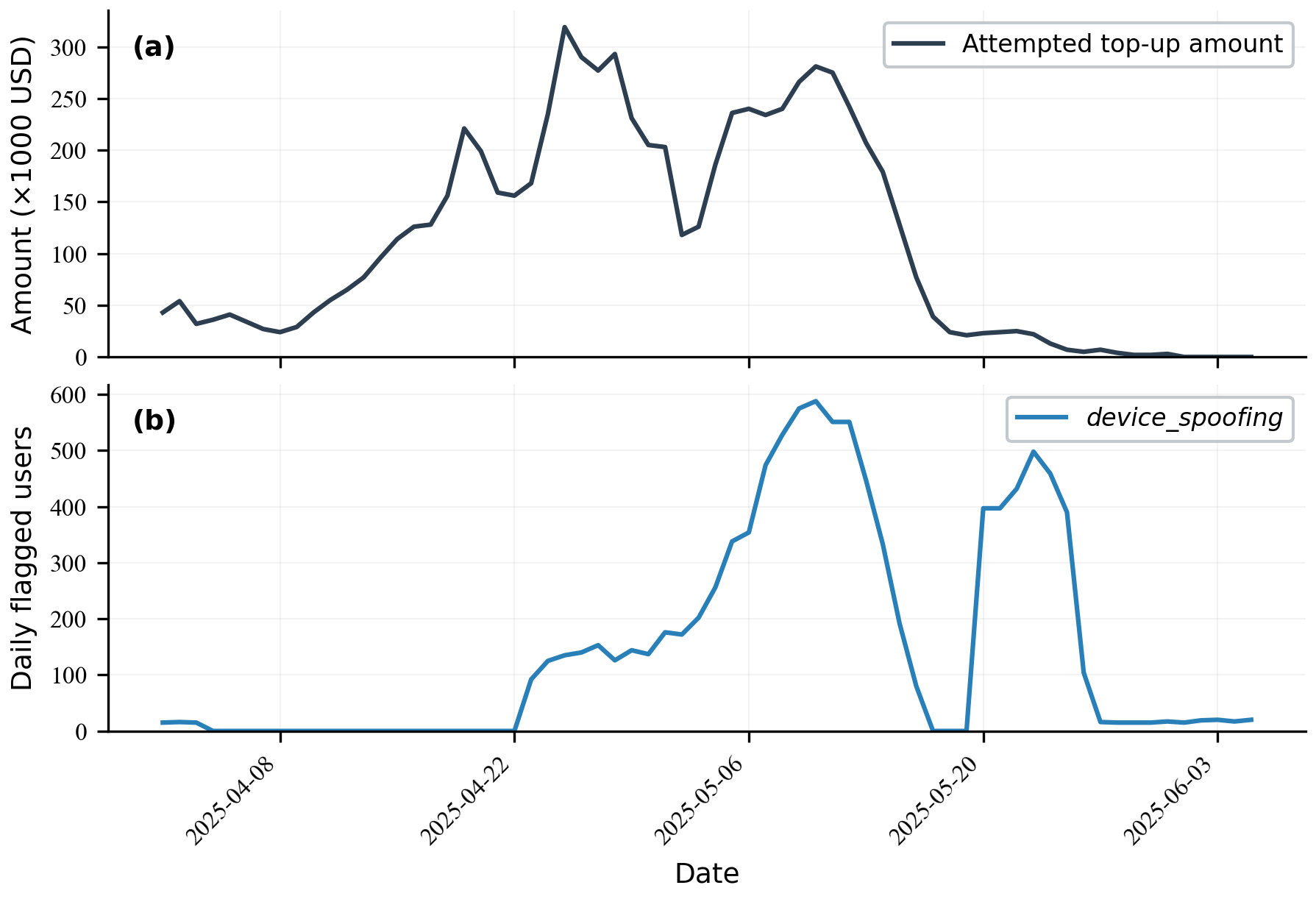}
  \caption{Case~2: Cross-modal detection of credit-card fraud. (a)~Daily attempted top-up volume from the payment system. (b)~Daily users flagged by \texttt{device\_spoofing} ($z{\geq}40.0$). The two curves exhibit correlated double-peak dynamics: the top-up surge in (a) precedes the detection surge in (b) by approximately 10--14 days---a natural delay since top-ups must occur before fraudulent spending triggers anomalous signal concentration at convergence merchants. Notably, \texttt{device\_spoofing} is a pre-existing, generic infrastructure-layer signal with zero business-logic connection to credit-card fraud, yet it tracks the attack timeline with high fidelity, demonstrating the framework's cross-modal detection capability.}
  \label{fig:case2-atu}
\end{figure}

\textbf{This finding reveals structural coupling at the infrastructure layer:} to control costs, attackers reuse the same underlying infrastructure (emulator clusters, scripted environments) across different business attack types, causing weak signals seemingly unrelated to the current attack to produce detectable abnormal density at convergence nodes.
\emph{It is worth emphasizing that \texttt{device\_spoofing} has zero business-logic connection to credit-card fraud, yet it covers 55.86\% of known fraudsters}---a coverage far exceeding random expectation, strongly confirming that cross-business-type infrastructure reuse is real and substantial.
This observation is consistent with predictions from the adversarial economics analysis (Section~\ref{sec:econ-model}): constrained by cost, attackers batch-script account manipulation and tend to leave homogeneous traces at the infrastructure layer (Section~\ref{sec:generalizability} further discusses the generalization implications).

\textbf{Attack-vector agnosticism evidence:}
Not all weak signals were ``activated'' in this incident.
Table~\ref{tab:case2-signals} shows the max $Z$-score of multiple weak signals during the incident; only signals related to the attack infrastructure exhibit pronounced anomalies:

\begin{table}[t]
  \centering
  \caption{Case~2: Selective activation of weak signals.}
  \label{tab:case2-signals}
  \small
  \begin{tabular}{@{}lccl@{}}
    \toprule
    \textbf{Weak signal} & \textbf{Max $z$ (incident)} & \textbf{Max $z$ (calm)} & \textbf{Activated?} \\
    \midrule
    \texttt{device\_spoofing} & $\sim$254 & $\sim$20--30 & \textbf{Yes} \\
    \texttt{payment\_failure} & $\sim$248 & $\sim$20--30 & \textbf{Yes} \\
    \texttt{foreign\_ip}      & $<$30     & $\sim$20--30 & No \\
    \bottomrule
  \end{tabular}
\end{table}

This \emph{selective activation} phenomenon demonstrates the framework's attack-vector agnosticism: the system need not know the current attack type in advance; structural aggregation automatically selects, from among multiple candidate weak signals, those relevant to the current attack.

\subsubsection{Quantitative Evaluation}\label{sec:case2-quant}

\textbf{Ground truth and labeling incompleteness:}
The ground truth for this case is the law-enforcement police report list (low hundreds of confirmed fraudsters).
This list stopped updating partway through the incident, while the system cumulatively flagged on the order of thousands of users---\emph{labels cover only about 10\% of system output}.
Therefore, Precision in Table~\ref{tab:case2-pr} is a nominal lower bound against the incomplete police list; a follow-up adjudication study (below) estimates actual precision far higher.

\textbf{Single-day snapshot detection performance} (the SCR denominator is the number of known fraudsters carrying the signal, 81):

\begin{table}[t]
  \centering
  \caption{Case~2: Single-day snapshot (device\_spoofing). Nominal Precision uses only the police-report list; adjudicated Precision at $z{=}40$ is reported in the text ($\geq 87\%$, 95\% CI).}
  \label{tab:case2-pr}
  \small
  \begin{tabular}{@{}cccccc@{}}
    \toprule
    $z$ & \textbf{HR Merch.} & \textbf{Flagged Users} & \textbf{Caught} & \textbf{Prec.} & \textbf{SCR} \\
    \midrule
    1.0  & 84 & 511 & 81 & 15.85\% & 100\% \\
    5.0  & 39 & 488 & 81 & 16.60\% & 100\% \\
    10.0 & 18 & 475 & 81 & 17.05\% & 100\% \\
    \textbf{40.0} & \textbf{7} & \textbf{466} & \textbf{81} & \textbf{17.38\%} & \textbf{100\%} \\
    \bottomrule
  \end{tabular}
\end{table}

Signal coverage $= 55.86\%$ (cross-modal signal coverage; this signal has no direct association with the attack's business logic; see Section~\ref{sec:cross-modal}).
Unconditional Recall $=$ SCR $\times$ signal coverage $= 100\% \times 55.86\% = 55.86\%$.

\textbf{Key observations:}

\begin{figure}[t]
  \centering
  \includegraphics[width=\columnwidth]{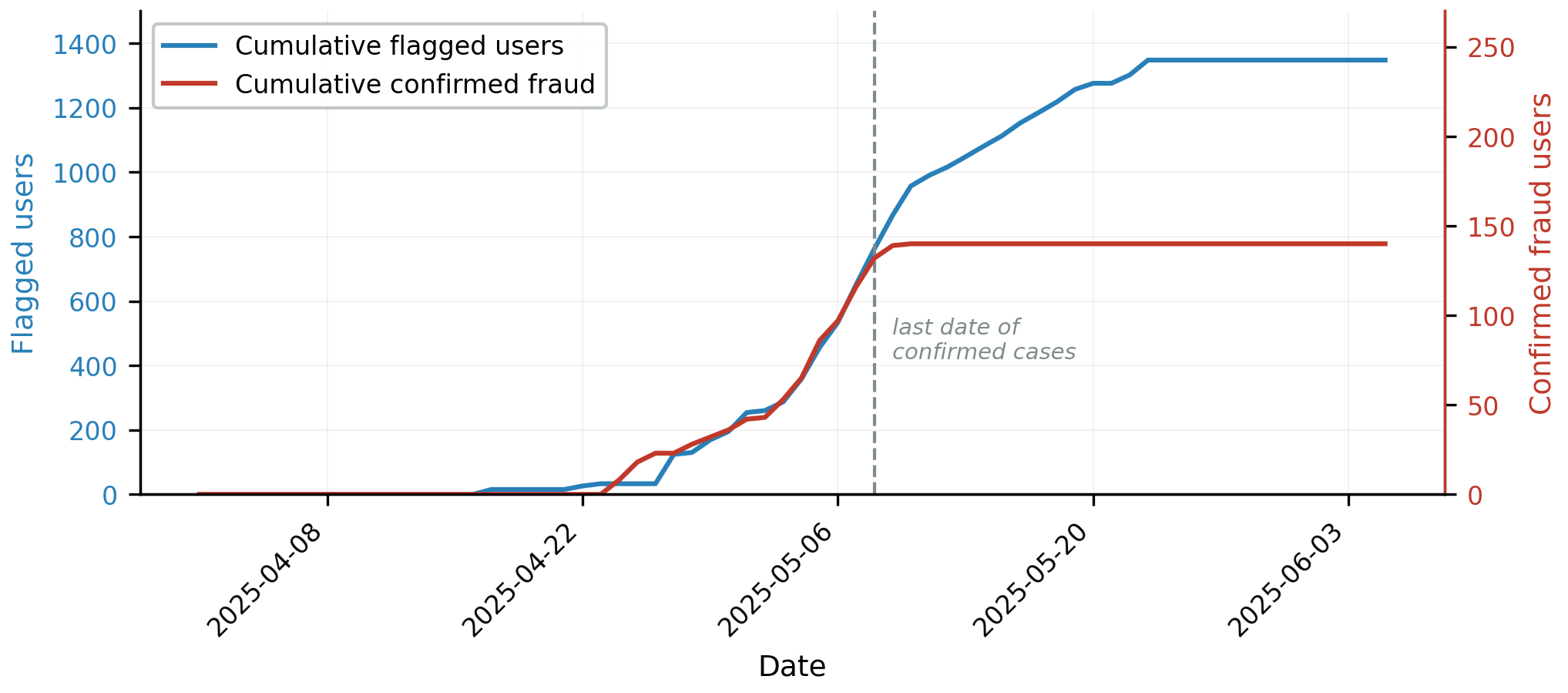}
  \caption{Case~2: Cumulative flagged users (blue, left axis) vs.\ cumulative confirmed fraud (red, right axis). The right axis is scaled so that both curves overlap during the co-growth phase (late April--early May), visually confirming proportional co-growth. The dashed vertical line marks the last date of confirmed cases; beyond it the confirmed-fraud curve plateaus while flagged users continue to accumulate---the resulting divergence reflects labeling incompleteness, not false positives. The absolute scale gap is ${\sim}10{:}1$ (${\sim}1{,}350$ flagged vs.\ ${\sim}140$ confirmed).}
  \label{fig:case2-cumulative}
\end{figure}

\begin{enumerate}[leftmargin=*]
  \item \textbf{Signal-conditioned Recall = 100\% (unchanged across all thresholds):} All 81 known fraudulent users carrying this signal are captured (81/81), indicating that fraud-associated merchants have extremely high $z$ values, making threshold choice irrelevant to recall (cross-case comparison in Section~\ref{sec:cross-case}).

  \item \textbf{Raising the threshold from $z{=}1.0$ to $z{=}40.0$ compresses high-risk merchants from 84 to 7 without losing any known-fraud captures.} Each merchant-level alert is accompanied by an associated suspicious-user list (466 users in total); the analysts' review unit is merchants rather than individual users, so seven alerts can be fully reviewed within minutes.

  \item \textbf{Nominal Precision (${\sim}17\%$) is a substantial underestimate of true precision at flag time.} Three lines of evidence, ordered by strength:

  \begin{itemize}[leftmargin=*]
    \item \textbf{Direct adjudication (primary):} We randomly sampled 20 of the 385 flagged-but-unlabeled users at the $z{=}40$ operating point and, ${\sim}15$ months later, reviewed independent enforcement outcomes. All 20 accounts had been deactivated; case remarks (bulk payment-fraud queues, chargeback appeals) were consistent with payment fraud rather than benign traffic. Treating these as true positives among the unlabeled yields an estimated unlabeled fraud rate $\hat{p}_u = 1.0$ (Wilson 95\% CI with finite-population correction $[0.85, 1.00]$), implying overall precision $P = (81 + \hat{p}_u \times 385)/466$ with a conservative 95\% lower bound of ${\geq}87\%$---far above the 17.38\% nominal figure. Many bans were recorded under \emph{subsequent} fraud cases, illustrating that absence from the police list at snapshot time reflects labeling and enforcement delay, not benign behavior at flag time.
    \item \textbf{Co-cessation (supporting):} Fig.~\ref{fig:case2-atu}(b) shows daily detection volume going to zero as fraudulent top-up volume (Fig.~\ref{fig:case2-atu}(a)) goes to zero; the cumulative flagged-user curve in Fig.~\ref{fig:case2-cumulative} likewise flattens quickly after the attack subsides, with no further new flags. If 83\% of apparent false positives were truly noise, that false-positive rate should persist after the attack ends. \emph{This argument assumes the composition of normal traffic at these merchants is stable across the attack cycle}---i.e., benign transaction patterns do not co-vary with the attack timeline.
    \item \textbf{Co-growth (supporting):} During the outbreak, cumulatively flagged users grew in near-proportional synchrony with known fraudulent users (Fig.~\ref{fig:case2-cumulative}), differing by approximately one order of magnitude---consistent with many unlabeled flags being latent true positives.
  \end{itemize}
\end{enumerate}

\textbf{Implications of labeling delay:}
This case vividly illustrates the labeling-delay dilemma described in Section~\ref{sec:related-supervised}: (1)~any supervised model relying on these labels will train on severely incomplete ground truth; (2)~even ex post evaluation will systematically underestimate Precision.
The proposed method requires no labels to operate and fundamentally sidesteps this dilemma.

\subsection{Cross-Case Analysis}\label{sec:cross-case}

The two cases validate the methodology's core claims from complementary angles (Table~\ref{tab:cross-case}).

\begin{table}[t]
  \centering
  \caption{Cross-case comparison.}
  \label{tab:cross-case}
  \small
  \begin{tabular}{@{}p{2.3cm}p{2.5cm}p{2.5cm}@{}}
    \toprule
    \textbf{Dimension} & \textbf{Case~1 (Promo)} & \textbf{Case~2 (Fraudulent Top-Up)} \\
    \midrule
    Fraud type & Non-injection (promo abuse) & Injection (credit-card fraud) \\
    Core weak signal & \texttt{use\_promo} (business layer) & \texttt{device\_spoofing} (infrastructure layer) \\
    Signal collection cost & Near zero (direct field read) & Low (device fingerprint matching) \\
    Signal--attack link & Direct association & \textbf{No direct association} (cross-modal) \\
    Rec.\ threshold & $z{=}40.0$ & $z{=}40.0$ \\
    Evaluation & Backtest + live deploy & Backtest \\
    Ground-truth quality & Manual review; relatively complete & Police report; \textbf{${\sim}10\%$ coverage} \\
    Max SCR & $\approx$100\% (99.94\%) & 100\% (all thresholds) \\
    Signal coverage & 99.82\% & 55.86\% \\
    Detection latency & 1 day & 1 day \\
    \bottomrule
  \end{tabular}
\end{table}

\textbf{Key observations:}

\begin{enumerate}[leftmargin=*]
  \item \textbf{Threshold consistency:} Across two independent cases spanning different markets, attack types, and weak signals, the same operating point $z{=}40.0$---chosen above each case's calm-period noise floor (max calm-period $z \approx 20$--$30$)---yields the highest precision of the thresholds we examined in both cases. The recall cost of that choice differs by case: it is zero in Case~2 (SCR stays at 100\%) but real in Case~1, where SCR falls from 99.94\% to 89.88\% (Table~\ref{tab:case1-pr}); operators trading recall for precision may prefer $z{=}10.0$ there. We present $z{=}40.0$ as an empirically transferable operating point, tuned relative to the observed noise floor, rather than a universal constant; that the same margin above the floor is usable in both scenarios is itself evidence of cross-scenario consistency.

  \item \textbf{Amplification of a minimal signal (Case~1):} \texttt{use\_promo} is the cheapest binary feature; used alone, global Precision is only 16\%. After amplification, at $z{=}10.0$, Precision reaches 91\% and unconditional Recall exceeds 99\% (${\sim}6{\times}$ amplification)---a canonical empirical demonstration of weak signal structural amplification.

  \item \textbf{Cross-modal effectiveness (Case~2):} An infrastructure-layer signal with zero business-logic connection to the attack (\texttt{device\_spoofing}) covers 55.86\% of known fraudsters, validating the framework's MO-agnostic property (see Section~\ref{sec:cross-modal}).

  \item \textbf{Max Signal-conditioned Recall approaches 100\% in both cases:} Under loose thresholds, the method captures almost all fraudsters carrying the signal (Market~A: 99.94\%, Market~B: 100\%). The large gap in Unconditional Recall between the two cases (Market~A 99.76\% vs Market~B 55.86\%) is entirely determined by signal coverage (99.82\% vs 55.86\%), not by differences in detection capability. Notably, in Market~B, SCR remains 100\% across all thresholds (the attack concentrates on very few merchants with extreme $z$ values), whereas in Market~A, SCR decreases from 99.94\% to 89.88\% at $z{=}40.0$, reflecting the Precision--Recall tradeoff when the attack is distributed across more medium-risk drivers.

  \item \textbf{Ground-truth quality contrast:} The two cases have very different ground-truth quality (manual review vs law-enforcement police report list), yet the method operates normally in both scenarios. Case~2's superficially low nominal Precision (${\sim}17\%$) is an empirical illustration of the labeling-delay dilemma (Section~\ref{sec:related-supervised}); follow-up adjudication estimates true precision $\geq 87\%$ (95\% CI)---the method itself does not degrade; it is the evaluation criterion (labels) that is incomplete.
\end{enumerate}

\subsection{Qualitative Comparison with Baseline Methods}\label{sec:baselines}

\begin{table}[t]
  \centering
  \caption{Qualitative comparison with baseline approaches.}
  \label{tab:baselines}
  \resizebox{\columnwidth}{!}{%
  \scriptsize
  \begin{tabular}{@{}lccccccc@{}}
    \toprule
    \textbf{Method} & \textbf{Prec.} & \textbf{Rec.} & \textbf{Compl.} & \textbf{Labels} & \textbf{Cold st.} & \textbf{Label-delay tol.} & \textbf{Interp.} \\
    \midrule
    Heuristic rules & Low & High & $O(n)$ & No & None & Unaffected & High \\
    Supervised ML~\cite{bolton2002statistical} & Med & Med & $O(n\!\log\!n)$ & Yes & Yes & \textbf{Severe} & Low \\
    GNN~\cite{dou2020enhancing,liu2021pick,lim2025fraudcengcl} & High & High & $O(n^2{+})$ & Yes & Yes & \textbf{Severe} & Low \\
    \textbf{Proposed} & \textbf{High} & \textbf{High} & $O(|E|)$ & \textbf{No} & \textbf{None} & \textbf{Unaffected} & \textbf{High} \\
    \bottomrule
  \end{tabular}%
  }
\end{table}

\emph{Note:} This table is a qualitative comparison.
We did not run a GNN baseline on the same dataset and therefore do not directly claim numerical superiority over supervised methods: training one would require exactly the delayed and incomplete labels whose absence motivates this work (Section~\ref{sec:case2-quant}), so a like-for-like comparison on these two incidents is not available to us.
A quantitative comparison against unsupervised dense-block detection and naive counting baselines, on public data, is reported in Section~\ref{sec:public-baselines}.

\textbf{Labeling-delay immunity:}
The proposed method is entirely independent of labeled data.
As empirically demonstrated in Section~\ref{sec:case2-quant}, reliable labels in fraud scenarios suffer from week- to month-level delays and are systematically incomplete; supervised methods can neither obtain training data in a timely manner nor reliably evaluate their own performance in this setting.
The proposed method is based on statistical aggregation and scoring rather than model training and can produce detection results on the \emph{same day} an attack occurs.

\textbf{Robustness:}
Combining the experimental evidence from both cases, the method demonstrates robustness along multiple dimensions:
(1)~\emph{across attack types}---effective in both economically distinct fraud modes, non-injection (Case~1) and injection (Case~2);
(2)~\emph{across weak signal sources}---both business-layer signals (\texttt{use\_promo}) and infrastructure-layer signals (\texttt{device\_spoofing}) achieve high-precision detection after amplification;
(3)~\emph{across label quality}---the method operates normally under both relatively complete labels (Case~1) and severely incomplete labels (Case~2, only ${\sim}10\%$);
(4)~\emph{adversarial robustness}---the detection target is a structural invariant dictated by economic constraints; evading it requires abandoning scale or absorbing the cost of dispersed cash-out, so full knowledge of the detection logic offers attackers no cheap countermeasure (Section~\ref{sec:limitations}).

\subsection{Quantitative Baseline Comparison on Public Data}\label{sec:public-baselines}

Since the production datasets cannot be released, we complement the case studies with a reproducible benchmark on public data, following the synthetic-attack protocol of FRAUDAR~\cite{hooi2016fraudar}.
The background graph is the SNAP Amazon Arts review graph~\cite{mcauley2013hidden} (24{,}070 users, 4{,}207 products, 27{,}655 edges), from which we draw five induced subgraphs of 20{,}000 users and 2{,}000 products by uniform random sampling under fixed seeds; backgrounds are drawn once and are not selected for attack detectability or for any method's performance.
Into each background we inject a block of 200 fraudulent users and 200 fraudulent products at edge densities $\{0.02, 0.04, 0.08\}$---a range chosen to span the undetectable-to-detectable regime rather than to reflect any measured density of real attacks---with three camouflage settings (none, random, biased), five injection seeds per configuration.
To emulate the framework's input, each user carries a binary weak signal calibrated to the Case~1 regime: it covers 80\% of fraudulent users, and benign users are assigned the signal at a rate targeting ${\sim}16\%$ global precision (realized precision $16.1\% \pm 0.7$~pp across trials, reported per trial in the released results).
Four methods rank products under a matched budget: the proposed EB-smoothed anomaly score, raw signal count, raw signal ratio, and FRAUDAR with its original logarithmic column weighting (validated against the reference example of~\cite{hooi2016fraudar}); FRAUDAR determines its own block size $K$, and all ranking methods flag their top-$K$ products.

\begin{table}[t]
  \centering
  \caption{Public-data benchmark, pooled over the detectable regime (injection density $\geq 0.04$; 150 trials per method). User-level metrics are computed over users attached to flagged products; recall is signal-conditioned.}
  \label{tab:public-baselines}
  \small
  \begin{tabular}{@{}lccc@{}}
    \toprule
    \textbf{Method} & \textbf{Product F1} & \textbf{User Prec.} & \textbf{User Rec.} \\
    \midrule
    \textbf{Proposed (EB $+$ score)} & \textbf{0.802} & \textbf{0.916} & \textbf{0.925} \\
    FRAUDAR~\cite{hooi2016fraudar} & 0.782 & 0.777 & 0.849 \\
    Raw signal count & 0.738 & 0.571 & 0.918 \\
    Raw signal ratio & 0.608 & 0.742 & 0.850 \\
    \bottomrule
  \end{tabular}
\end{table}

Table~\ref{tab:public-baselines} pools the detectable regime.
Two caveats qualify these numbers.
First, the attacks are synthetic injections into a real background, not labeled real fraud; the benchmark measures ranking behavior under a controlled convergence structure, not end-to-end detection of organic abuse.
Second, the proposed method and the naive baselines consume the weak signal, whereas FRAUDAR is purely structural; the comparison therefore shows that aggregating a cheap weak signal onto convergence nodes is competitive with dense-block search, not that the scoring function dominates FRAUDAR under equal information.
At density $0.02$ all four methods fail (F1 near zero), which empirically locates the detectability limit of this background/attack scale.

\textbf{Scoring function versus structural aggregation:}
The benchmark also clarifies \emph{where} the framework's precision comes from.
When an attack is extremely concentrated, any monotone aggregate of the weak signal will tend to surface the same few convergence nodes, and we observed exactly that: in a matched-budget re-analysis of the two production incidents, ranking by raw signal count selected the same node set as the anomaly score at $z{=}40$.
Two caveats bound what this observation supports.
Exact agreement is confined to the extreme threshold: at lower operating points the two rankings select node sets that overlap substantially (78--98\%) but not completely, although the resulting precision differs by well under one point.
In Case~2 the matched budget at $z \geq 5$ is a single merchant, so agreement there is close to vacuous---any ranking that puts the obvious node first will match.
Because the re-analysis is run on reconstructed snapshots, it should be read as evidence about ranking behaviour on a comparable graph rather than as a restatement of the case-study numbers of Sections~\ref{sec:case1-quant} and~\ref{sec:case2-quant}.
The claim we therefore make is narrow: \emph{in these two incidents, at an operating point chosen far above the noise floor, the scoring layer conferred no advantage over counting that this analysis could detect}.
We do not claim that the scoring function contributes nothing to precision in general---the public benchmark reported above is a direct counterexample---nor that the concentrated regime is the one deployments will always face.
The scoring layer earns its keep at the margins and in operations: on the public benchmark, where injected blocks compete with high-volume benign products, raw-count ranking incurs a large user-level false-positive cost (precision 0.571 vs.\ 0.916) because it systematically favors large nodes regardless of signal proportion, while EB smoothing suppresses both large benign nodes and small-sample noise.
Moreover, the calibrated score is what makes the calm-period noise-floor analysis (Section~\ref{sec:method-exp}) and the fixed cross-scenario operating point (Section~\ref{sec:cross-case}) meaningful: raw counts are not comparable across nodes of different sizes or across days of different traffic volumes.
Read together, the two settings suggest that the scoring layer's contribution is a function of how concentrated the attack is: it was undetectable in the extreme regime of our production operating point and large in the more competitive regime of the public benchmark.
We accordingly present it as a calibration and false-positive-control mechanism whose value grows as an attack becomes less extreme---not as the sole source of the reported precision, and not as dispensable.
That the structural core degrades gracefully to counting when the attack is blatant is, in our view, an argument for the framework's deployability rather than against it.

% ============================================================
\section{Discussion}\label{sec:discussion}

\subsection{Applicability and Limitations}\label{sec:limitations}

The method is designed for \emph{burst incidents at large scale} and relies on weak signals forming sufficient statistical aggregation at convergence nodes; its ability to detect low-frequency individual fraud is limited.
For online services, however, \emph{burst incidents at large scale are the core threat that truly merits resource investment}---a single large-scale burst incident can cause losses on the order of millions of US dollars, whereas the impact of small-scale dispersed fraud is comparatively limited.

\textbf{Statement of limitations:}
\begin{itemize}[leftmargin=*]
  \item For low-frequency, dispersed individual fraud, the concentration of weak signals at convergence nodes may be insufficient to lift the anomaly score above the operating threshold.
  \item If attackers have sufficient resources to maintain a large number of dispersed cash-out nodes (violating the Low Cost constraint of the Trilemma), the effectiveness of the method is reduced.
\end{itemize}

\textbf{Defense economics and the ``open-hand'' property:}
The above limitations themselves constitute a defensive effect.
We distinguish two claims here.
First, the \emph{structural invariant} (centralized cash-out) is economically hard to escape: it is a corollary of the Fraudster's Trilemma, and escaping it requires the attacker to abandon scale or bear the cost of dispersed cash-out---both erode the economics that motivate the attack in the first place.
Second, and more modestly, the \emph{specific detector} built on this invariant retains value under full disclosure: unlike rules or models that rely on secrecy, its detection logic can be published (``open-hand''), because circumventing it pushes the attacker toward the costly operating modes above rather than a cheap tactical tweak.
Evasion is not impossible---an informed attacker may, for example, dilute the weak-signal proportion at a convergence node by routing benign traffic through it, or rotate away from the specific weak signals being aggregated---but each such countermeasure consumes resources and degrades the attack's return on investment.
From a game-theoretic perspective, raising the cost of evasion to the point where the attack becomes economically unattractive itself deters attack motivation.
This property reduces---though does not eliminate---the need for continual updating in long-term adversarial settings, in contrast to rule systems whose effectiveness depends on ongoing tuning.

\subsection{Generalizability}\label{sec:generalizability}

The conditions under which the framework applies are the presence of a many-to-few convergence structure and the availability of weak signals.
The cross-modal detection results of Case~2 (Section~\ref{sec:cross-modal}) have empirically demonstrated the MO-agnostic nature of the framework: the framework relies primarily on the structural invariant and is insensitive to the specific source of weak signals.
Our direct evidence, however, covers two incident types in two adjacent markets on a single platform; we therefore state the following as \emph{hypotheses} rather than validated claims:

\begin{itemize}[leftmargin=*]
  \item \textbf{Across fraud types:} Payment fraud, promotional abuse, and account takeover share the centralized cash-out structure in principle, and reputation-manipulation settings (Sybil accounts $\to$ target merchants/products) likewise possess a many-to-few convergence topology. We expect the method to extend to these settings, but none of them is experimentally validated in this paper.
  \item \textbf{Across platforms:} Platforms with convergence of funds or behavior (e-commerce, payments, social) satisfy the framework's structural precondition; validating the transfer across platforms remains future work.
\end{itemize}

\subsection{Ethics and Data Handling}\label{sec:ethics}

This work analyzes transaction records and device-environment signals that the platform collects in the course of its normal fraud-prevention operations, processed under the platform's terms of service and privacy policy.
All data access and analysis were performed by authorized employees within the platform's production environment, subject to the platform's data-retention and access-control policies; no raw data left that environment.
The paper reports only aggregate, de-identified statistics: user and merchant identifiers are omitted, and incident-level figures (monetary amounts, account and merchant counts) are generalized to orders of magnitude.
The Case~2 ground truth derived from a law-enforcement report was used solely as evaluation labels and is not redistributed.
External publication of this work was reviewed and approved through the platform's internal publication-review process, including communications and management review.

Regarding potential harm from false positives: the system's output is a node-level alert with an associated user list, used for \emph{investigation prioritization} rather than automatic punishment---enforcement actions (e.g., account blocking) are taken only after human analyst review (Section~\ref{sec:case1-quant}), and affected users may contest actions through the platform's standard appeal channels.
The high precision at the recommended operating point further limits collateral impact.

% ============================================================
\section{Conclusion}\label{sec:conclusion}

This paper proposes the \textbf{Fraudster's Trilemma} theory and, building on it, designs a structural amplification method for weak signals to detect organized abuse on large online platforms.

\textbf{Core logic chain:}
\begin{gather*}
  \textbf{Fraudster's Trilemma} \;\xrightarrow{\text{corollary}} \\
  \text{structural invariant (centralized cash-out)} \;\xrightarrow{\text{exploit}} \\
  \text{structural amplification of weak signals} \;\xrightarrow{\text{achieve}} \\
  \text{High-precision fraud detection}
\end{gather*}

\textbf{Summary of main contributions:}
\begin{enumerate}[leftmargin=*]
  \item We propose the \textbf{Fraudster's Trilemma}: organized attackers cannot simultaneously achieve scale, low cost, and dispersed cash-out. This theory reveals the structural invariant of organized fraud---centralized cash-out---and provides a theoretical basis for detection methods.

  \item We show that structural aggregation of weak signals at convergence nodes can amplify low-precision signals into high-precision decisions---inexpensive weak signals, after amplification, achieve high-precision detection in two distinct fraud incidents, and infrastructure-layer weak signals successfully detect payment-layer attacks without business-logic linkage (cross-modal effectiveness).

  \item The method is label-free, nearly parameter-free, free of cold start, of linear complexity, and white-box interpretable; it has been validated as effective in real production and can produce detection results on the \emph{same day} the attack occurs, fundamentally avoiding the labeling-delay dilemma in fraud settings. Moreover, because the detection target is a structural invariant determined by economic constraints, the method retains its value under full disclosure---evading it forces attackers to abandon scale or bear the high cost of dispersed cash-out, substantially raising the cost of evasion.
\end{enumerate}

% ============================================================
\bibliographystyle{IEEEtran}
\bibliography{paper}

\end{document}